\documentclass[aps,pre,reprint,letterpaper,showpacs,amsmath,amssymb]{revtex4-1}
\usepackage[absolute]{textpos}
\begin{document}
\begin{textblock*}{8.5in}(0.1in,0.25in)
\begin{center}
PHYSICAL REVIEW E \textbf{85}, 061120 (2012)
\end{center}
\end{textblock*}
\begin{textblock*}{2.5in}(5.6in,10.5in)
\copyright2012 American Physical Society
\end{textblock*}
\title{Hierarchical maximum entropy principle
for generalized superstatistical systems and\\
Bose-Einstein condensation of light}
\received{21 February 2012}
\published{20 June 2012}
\author{Denis Nikolaevich \surname{Sob'yanin}}
\email{sobyanin@lpi.ru}
\affiliation{Tamm Department of Theoretical Physics,\\Lebedev Physical
Institute, Russian Academy of Sciences,\\Leninskii Prospekt 53, Moscow, 119991
Russia}
\begin{abstract}
A principle of hierarchical entropy maximization is proposed for generalized
superstatistical systems, which are characterized by the existence of three
levels of dynamics. If a generalized superstatistical system comprises a set of
superstatistical subsystems, each made up of a set of cells, then the
Boltzmann-Gibbs-Shannon entropy should be maximized first for each cell,
second for each subsystem, and finally for the whole system. Hierarchical
entropy maximization naturally reflects the sufficient time-scale separation
between different dynamical levels and allows one to find the distribution of
both the intensive parameter and the control parameter for the corresponding
superstatistics. The hierarchical maximum entropy principle is applied to
fluctuations of the photon Bose-Einstein condensate in a dye microcavity. This
principle provides an alternative to the master equation approach recently
applied to this problem. The possibility of constructing generalized
superstatistics based on a statistics different from the Boltzmann-Gibbs
statistics is pointed out.
\end{abstract}
\pacs{05.20.--y, 05.30.--d, 05.70.--a, 67.85.--d}
\maketitle
\section{Introduction}

Superstatistics represents a statistics of canonical statistics and allows one
to consider stationary states of nonequilibrium systems with fluctuations of an
intensive parameter~$\beta$ \cite{BeckCohen2003}. Though usually considered as
an inverse temperature, $\beta$ can be interpreted in a more general way
\cite{BeckCohenSwinney2005,Beck2009}. A superstatistical system comprises a set
of subsystems, or cells, each having the Gibbs canonical distribution determined
by~$\beta$. An essential feature of the superstatistical system is sufficient
spatiotemporal scale separation, so that $\beta$ fluctuates on a much larger
time scale than the typical relaxation time of the local dynamics in a cell.
Superstatistics can be given a basis by the theory of hyperensembles
\cite{Crooks2007,Abe2009}.

The distribution of~$\beta$ can be considered as a function of some additional
control parameters \cite{StraetenBeck2008}. However, in  ordinary
superstatistics, the intensive parameter fluctuates, but the control parameters
are constant. Considering the control parameter fluctuations has led very
recently to the generalization of superstatistics---``statistics of
superstatistics,'' or ``generalized superstatistics'' \cite{Sobyanin2011}.
Generalized superstatistics is the statistics of generalized superstatistical
systems. A generalized superstatistical system comprises a set of nonequilibrium
superstatistical subsystems and can be associated with a generalized
hyperensemble, an ensemble of hyperensembles. Compared with an ordinary
superstatistical system, a generalized superstatistical system is characterized
by the existence of the third, upper level of dynamics in addition to the two
levels of dynamics existing in each superstatistical subsystem. This is
reflected in the existence of a fluctuating vector control parameter on which
both the intensive parameter distribution and the density of energy states
depend. Significantly, generalized superstatistics can be used for nonstationary
nonequilibrium systems. It was applied to branching processes and pair
production in a neutron star magnetosphere \cite{Sobyanin2011}.

The main problem of generalized superstatistics is the determination of the
intensive parameter distribution, characterizing the superstatistical dynamics
in each subsystem, and the control parameter distribution, characterizing the
dynamics of the system as a whole. The aim of this paper is to develop the
maximum entropy principle that can be used to solve the above problem.

The paper is organized as follows: In Sec.~II the hierarchical maximum entropy
principle for generalized superstatistical systems is formulated and the
canonical, intensive parameter, and control parameter distributions are
consecutively determined. In Sec.~III this principle is applied to Bose-Einstein
condensation of light and fluctuations of the number of ground-mode photons are
considered. In Sec.~IV the main conclusions are given.

\section{Hierarchical maximum entropy}

A generalized superstatistical system is conveniently thought of as a set of
superstatistical subsystems, each in turn made up of a set of cells. There are
three levels of dynamics in this system: the first, lower level of fast dynamics
in a cell, the second, middle level of superstatistical dynamics in a subsystem,
and the third, upper level of global dynamics in the whole system. The levels
are arranged in increasing order of dynamical time scale so that the shortest
time scale corresponds to the lower level. The local
dynamics in a cell is characterized by an energy $E$, the superstatistical
dynamics in a subsystem is characterized by an intensive parameter $\beta$, and
the global dynamics in the whole system is characterized by a control parameter
$\xi$, which may be a multidimensional vector.

The system hierarchy is formed as a result of the sufficient time-scale
separation between different levels of dynamics. This allows us to formulate the
maximum entropy principle for the generalized superstatistical system as a
principle of hierarchical entropy maximization. More specifically, the entropy
should be maximized first for each cell, second for each subsystem, and
finally for the whole system.

\subsection{Local dynamics}

Though the existence of the Gibbs canonical distribution at the lower dynamical
level is postulated in superstatistics, it is reasonable to explicitly obtain
this distribution from the maximum entropy
principle. This trivial derivation will allow us to readily observe an analogy
between the dynamics at different hierarchical levels of a generalized
superstatistical system.

Choose a superstatistical subsystem of the generalized superstatistical system.
A fixed value of the control parameter $\xi$ corresponds to this subsystem, but
the intensive parameter $\beta$ may still fluctuate. Choosing the subsystem also
fixes the density of energy states:
\begin{equation}
\label{densityOfStates}
g(E|\xi)=\frac{\partial\Gamma(E|\xi)}{\partial E},
\end{equation}
where $\Gamma(E|\xi)$ is the number of states with energy less than $E$. In
integrals with $d\Gamma(E|\xi)$, integration over $E$ will be performed,
$d\Gamma(E|\xi)=g(E|\xi)dE$.

To consider the local dynamics, choose a cell of the subsystem. Then $\beta$
also becomes fixed, but the energy $E$ is not fixed and is characterized by a
probability distribution $\rho(E|\beta,\xi)$. To find the distribution
maximizing the Boltzmann-Gibbs-Shannon entropy
\[
S[E](\beta|\xi)=-\int\rho(E|\beta,\xi)\ln\rho(E|\beta,\xi)d\Gamma(E|\xi)
\]
under the normalization condition $N[E](\beta|\xi)=1$ and the mean energy
constraint $U[E](\beta|\xi)=U(\beta|\xi)$, where
\begin{eqnarray*}
N[E](\beta|\xi)&=&\int\rho(E|\beta,\xi)d\Gamma(E|\xi),\\
U[E](\beta|\xi)&=&\int E\rho(E|\beta,\xi)d\Gamma(E|\xi),
\end{eqnarray*}
we should consider the condition of zero variation, $\delta L_1=0$, for the
Lagrange function
\[
L_1(\nu_1,\beta,\xi)=S[E](\beta|\xi)-(\nu_1-1)N[E](\beta|\xi)
-\beta U[E](\beta|\xi).
\]
Then we arrive at the Gibbs canonical distribution
\[
\rho_G(E|\beta,\xi)=\frac{e^{-\beta E}}{Z(\beta|\xi)},
\]
where
\begin{equation}
\label{partitionFunction}
Z(\beta|\xi)=\int e^{-\beta E}d\Gamma(E|\xi)
\end{equation}
is the partition function. The entropy is
\begin{equation}
\label{entropyInEachCell}
S[E](\beta|\xi)=\nu_1(\beta|\xi)+\beta U(\beta|\xi),
\end{equation}
where the mean energy
\begin{equation}
\label{meanEnergyInACell}
U(\beta|\xi)=-\frac{\partial\nu_1(\beta|\xi)}{\partial\beta}
\end{equation}
is expressed via the Massieu function
\begin{equation}
\label{nu1}
\nu_1(\beta|\xi)=\ln Z(\beta|\xi).
\end{equation}

\subsection{Superstatistical dynamics}

Now consider the superstatistical dynamics of the chosen subsystem. This
dynamics is characterized by the fluctuating intensive parameter $\beta$ that
determines the properties of cells of the subsystem. To find the intensive
parameter distribution $f(\beta|\xi)$, we should maximize the entropy of the
joint probability distribution of~$E$ and~$\beta$, given~$\xi$. It is written as
\cite{AbeBeckCohen2007,Abe2009}
\begin{equation}
\label{subsystemEntropy}
S[E,\beta](\xi)=S[\beta](\xi)+\int S[E](\beta|\xi)f(\beta|\xi)d\beta
\end{equation}
where
\begin{equation}
\label{betaEntropy}
S[\beta](\xi)=-\int f(\beta|\xi)\ln f(\beta|\xi)d\beta
\end{equation}
is the entropy associated with
$f(\beta|\xi)$, and $S[E](\beta|\xi)$ is given by Eq.~\eqref{entropyInEachCell}.
The normalization condition for $f(\beta|\xi)$ is $N[\beta](\xi)=1$, where
\[
N[\beta](\xi)=\int f(\beta|\xi)d\beta.
\]
In addition, we may impose a set of $n$ constraints given by an $n$-dimensional
vector equality
\begin{equation}
\label{middleLevelConstraintsEqualToZero}
M[\beta](\xi)=M(\xi),
\end{equation}
where
\begin{equation}
\label{middleLevelConstraints}
M[\beta](\xi)=\int m(\beta|\xi)f(\beta|\xi)d\beta,
\end{equation}
and $m(\beta|\xi)=[m_1(\beta|\xi),\ldots,m_n(\beta|\xi)]$ and
$M(\xi)=[M_1(\xi),\ldots,M_n(\xi)]$ are $n$-dimensional vectors specifying,
respectively, the form and values of the constraints. Each $M_i(\xi)$ is the
mean of $m_i(\beta|\xi)$ over the fluctuating $\beta$, given~$\xi$. We consider
$M[\beta](\xi)$ as some general constraint vector, but it may be composed of the
constraints used in ordinary superstatistics, e.g., the mean values of energy,
entropy, square of entropy, energy divided by temperature, or logarithm of the
partition function \cite{Crooks2007,Naudts2007,StraetenBeck2008,Abe2010}.

Also define an $n$-dimensional vector Lagrange multiplier
$\mu=(\mu_1,\ldots,\mu_n)$, where each $\mu_i$ is the Lagrange multiplier
corresponding to the constraint $M_i[\beta](\xi)=M_i(\xi)$. We then have the
following Lagrange function:
\[
L_2(\nu_2,\mu,\xi)=
S[E,\beta](\xi)-(\nu_2-1)N[\beta](\xi)-\mu\cdot M[\beta](\xi).
\]
By $a\cdot b=\sum a_i b_i$ we denote the scalar product of some vectors $a$
and~$b$. The condition $\delta L_2=0$ yields the intensive parameter
distribution
\begin{equation}
\label{superstatisticalDistribution}
\tilde{f}(\beta|\mu,\xi)=
\frac{Z(\beta|\xi)}{\tilde{Y}(\mu,\xi)}\exp[-\mu\cdot m(\beta|\xi)+\beta
U(\beta|\xi)],
\end{equation}
where the partition function
\begin{equation}
\label{YMuXi}
\tilde{Y}(\mu,\xi)=\int Z(\beta|\xi)\exp[-\mu\cdot m(\beta|\xi)+\beta
U(\beta|\xi)]d\beta
\end{equation}
is determined from the normalization condition for~$\tilde{f}(\beta|\mu,\xi)$.

Note that $\tilde{f}(\beta|\mu,\xi)$ and $\tilde{Y}(\mu,\xi)$ still depend on
the Lagrange multiplier~$\mu$. The implicit dependence of~$\mu$ on the control
parameter~$\xi$,
\begin{equation}
\label{LagrangeMultiplierOnControlParameter}
\mu=\mu(\xi),
\end{equation}
is determined from
\begin{equation}
\label{MXi}
M(\xi)=-\frac{\partial\tilde{\nu}_2(\mu,\xi)}{\partial\mu},
\end{equation}
where
\begin{equation}
\label{nu2}
\tilde{\nu}_2(\mu,\xi)=\ln\tilde{Y}(\mu,\xi)
\end{equation}
is the Massieu function and
$\partial/\partial\mu=(\partial/\partial\mu_1,\ldots,\partial/\partial\mu_n)$ is
the $n$-dimensional gradient operator. Equations \eqref{MXi} and~\eqref{nu2} are
analogous to Eqs.~\eqref{meanEnergyInACell} and~\eqref{nu1}, respectively. Thus,
given the constraints~\eqref{middleLevelConstraintsEqualToZero}, the intensive
parameter distribution~\eqref{superstatisticalDistribution}, partition
function~\eqref{YMuXi}, and Massieu function \eqref{nu2} depend only on~$\beta$
and~$\xi$:
\begin{equation}
\label{finalSuperstatisticalDistribution}
f(\beta|\xi)=\tilde{f}(\beta|\mu(\xi),\xi),
\end{equation}
\begin{equation}
\label{Yxi}
Y(\xi)=\tilde{Y}(\mu(\xi),\xi),
\qquad
\nu_2(\xi)=\tilde{\nu}_2(\mu(\xi),\xi).
\end{equation}

We may either first set the constraint vector $M(\xi)$ and then find $\mu(\xi)$
from the maximum entropy principle, or vice versa. This is in full analogy with
the case of the dynamics in a cell, when we may first set the mean energy
$U(\beta)$ and then find the corresponding intensive parameter $\beta$, or set
$\beta$ and then find $U(\beta)$, which is more common. Incidentally, this duality
allows one to alternatively formulate superstatistics by introducing the
fluctuations of~$U(\beta)$ instead of those of~$\beta$ \cite{Bercher2008}. Note
that the control parameter $\xi$ has a more general nature than $\beta$, since
$\beta$ is exactly a Lagrange multiplier, while $\xi$, though controlling the
Lagrange multiplier~$\mu$, may not coincide with~$\mu$. The analogy between
$\beta$ and $\xi$ will be complete if we choose $\mu(\xi)=\xi$.

It follows from Eqs.~\eqref{entropyInEachCell},
\eqref{nu1}--\eqref{superstatisticalDistribution},
\eqref{LagrangeMultiplierOnControlParameter},
and \eqref{nu2}--\eqref{Yxi} that the entropy
associated with the superstatistical subsystem is
\begin{equation}
\label{finalSubsystemEntropy}
S[E,\beta](\xi)=\nu_2(\xi)+\mu(\xi)\cdot M(\xi).
\end{equation}
It is analogous to Eq.~\eqref{entropyInEachCell}.

Thus, the intensive parameter distribution for the superstatistical subsystem is
given by Eq.~\eqref{finalSuperstatisticalDistribution}. The superstatistical
distribution
\[
\rho(E|\xi)=\int\rho_G(E|\beta,\xi)f(\beta|\xi)d\beta
\]
has the form
\begin{equation}
\label{ultimateSuperstatisticalDistribution}
\rho(E|\xi)=\frac{1}{Y(\xi)}\int\exp\{-\beta[E-U(\beta|\xi)]-\mu(\xi)\cdot
m(\beta|\xi)\}d\beta,
\end{equation}
with the normalization condition $\int\rho(E|\xi)d\Gamma(E|\xi)=1$.

Ordinary superstatistics is a special case of generalized superstatistics: an
ordinary superstatistical system is a generalized superstatistical system
without fluctuations of the control parameter $\xi$. Therefore, we can easily
obtain the intensive parameter distribution $f=f(\beta|\mu)$ for this system by
formally removing $\xi$ from Eq.~\eqref{superstatisticalDistribution} and from
subsidiary Eqs.~\eqref{densityOfStates}, \eqref{partitionFunction},
\eqref{meanEnergyInACell}, \eqref{nu1},
\eqref{middleLevelConstraintsEqualToZero}, \eqref{middleLevelConstraints},
\eqref{YMuXi}, \eqref{MXi}, and \eqref{nu2}. It is consistent with the
distributions obtained earlier \cite{AbeBeckCohen2007,StraetenBeck2008,Abe2010}.

\subsection{Global dynamics}

Consider the third level of dynamics. We should find the probability
distribution $c(\xi)$ of the fluctuating control parameter $\xi$. This
distribution is normalized,
$N[\xi]=1$, where
\[
N[\xi]=\int c(\xi)d\xi.
\]
The entropy of the joint probability distribution of~$E$, $\beta$, and~$\xi$ is
determined by analogy with the entropy associated with a superstatistical
subsystem [cf.~Eq.~\eqref{subsystemEntropy}]:
\begin{equation}
\label{totalEntropy}
S[E,\beta,\xi]=S[\xi]+\int S[E,\beta](\xi)c(\xi)d\xi,
\end{equation}
where
\begin{equation}
\label{controlParameterDistributionEntropy}
S[\xi]=-\int c(\xi)\ln c(\xi)d\xi
\end{equation}
is the entropy associated with the control parameter distribution $c(\xi)$, and
$S[E,\beta](\xi)$ is given by Eq.~\eqref{finalSubsystemEntropy}.
We may impose a set of $m$ additional constraints by analogy with
Eqs.~\eqref{middleLevelConstraintsEqualToZero}
and~\eqref{middleLevelConstraints}:
\begin{equation}
\label{thirdLevelConstraintsVector}
K[\xi]=K,
\end{equation}
where
\[
K[\xi]=\int k(\xi)c(\xi)d\xi,
\]
and $k(\xi)=[k_1(\xi),\ldots,k_m(\xi)]$ and $K=(K_1,\ldots,K_m)$ are
$m$-dimensional vectors specifying, respectively, the form and values of the
constraints. Each $K_i$ is the mean of $k_i(\xi)$ over the fluctuating~$\xi$.

The Lagrange function is
\[
L_3(\nu_3,\kappa)=S[E,\beta,\xi]-(\nu_3-1)N[\xi]-\kappa\cdot K[\xi],
\]
where we have defined an $m$-dimensional vector Lagrange multiplier
$\kappa=(\kappa_1,\ldots,\kappa_m)$, where each $\kappa_i$ is the Lagrange
multiplier corresponding to the constraint $K_i[\xi]=K_i$. The condition
$\delta L_3=0$ yields the control parameter distribution
\begin{equation}
\label{controlParameterDistribution}
c(\xi,\kappa)=\frac{Y(\xi)}{X(\kappa)}\exp[-\kappa\cdot k(\xi)+\mu(\xi)\cdot
M(\xi)],
\end{equation}
where the partition function is
\[
X(\kappa)=\int Y(\xi)\exp[-\kappa\cdot k(\xi)+\mu(\xi)\cdot M(\xi)]d\xi,
\]
and $Y(\xi)$ is defined by Eq.~\eqref{Yxi}. By analogy with Eq.~\eqref{MXi}, we
can rewrite the constraints \eqref{thirdLevelConstraintsVector} as follows:
\begin{equation}
\label{kappa}
K=-\frac{\partial\nu_3(\kappa)}{\partial\kappa},
\end{equation}
where
\[
\nu_3(\kappa)=\ln X(\kappa)
\]
is the Massieu function, and
$\partial/\partial\kappa=(\partial/\partial\kappa_1,\ldots,
\partial/\partial\kappa_m)$ is the $m$-dimensional gradient operator.
It remains to find the entropy \eqref{totalEntropy} at the maximum point
[cf.~Eqs.~\eqref{entropyInEachCell} and~\eqref{finalSubsystemEntropy}]:
\[
S[E,\beta,\xi]=\nu_3(\kappa)+\kappa\cdot K.
\]

Thus, the intensive parameter distribution $c(\xi)\equiv c(\xi,\kappa)$ is given
by Eq.~\eqref{controlParameterDistribution}, with the Lagrange multiplier
$\kappa$ determined from Eq.~\eqref{kappa}. By
Eqs.~\eqref{ultimateSuperstatisticalDistribution}
and~\eqref{controlParameterDistribution}, we get that the generalized
superstatistical distribution
\[
\sigma(E)=\int\rho(E|\xi)g(E|\xi)c(\xi)d\xi
\]
has the form
\begin{eqnarray*}
\sigma(E)&=&\frac{1}{X(\kappa)}\int\exp\{-\beta[E-U(\beta|\xi)]\\
& &-\mu(\xi)\cdot[m(\beta|\xi)-M(\xi)]
-\kappa\cdot k(\xi)\}g(E|\xi)d\beta d\xi,
\end{eqnarray*}
with the normalization condition $\int\sigma(E)dE=1$.

\section{Bose-Einstein condensation of light}

Recently, thermalization of light in a dye microcavity has been observed
\cite{KlaersVewingerWeitz2010}. In this experiment, photons are confined in a
curved-mirror optical microresonator filled with a dye solution. In the
microresonator, absorption and reemission of photons by dye molecules results
in thermalization of the photon gas. Since the free spectral range of the
microresonator is comparable to the spectral width of the dye, the emission of
photons with a fixed longitudinal number dominates. Therefore, the photon gas is
effectively two dimensional, and thermalization of transverse photon states
occurs. Moreover, Bose-Einstein condensation (BEC) of light has been
experimentally observed in the described system
\cite{KlaersEtal2010,KlaersEtal2011}. This reflects the fact that a
two-dimensional harmonically trapped ideal gas of massive bosons can undergo BEC
\cite{BagnatoKleppner1991,Mullin1997,WeissWilkens1997,
KocharovskyEtal2006,PitaevskiiStringari2003}.
In the case of the light BEC, the curvature of the
mirrors provides a nonvanishing effective photon mass and at the same time
induces a harmonic trapping potential for photons.

The problem of thermalization and fluctuations of the photon Bose-Einstein
condensate has been considered very recently in
Ref.~\cite{KlaersEtal2012}. The condensate exchanges excitations with a
reservoir consisting of $M$ dye molecules. The authors assume that the
ground-state photon mode is coupled to the electronic transitions of a given
number of dye molecules. This means that the sum $X$ of the number of
ground-mode photons, $n$, and that of excited dye molecules, $X-n$, is constant.
To analyze this system, the authors use the master equation approach.

Note that if we are interested in
the behavior of the fluctuating photon BEC after thermalization has occurred, we
can obtain the corresponding probability distribution merely using the
thermodynamic consideration. The population of the electronic states of dye
molecules is quickly thermalized, with the characteristic time ${\sim}1$~ps at
room temperature
(see Refs.~\cite{Shank1975,SchaeferWillis1976,
HaasRotter1991,Schaefer1990,Lakowicz2006} for details).
Since the typical fluorescence lifetime is
${\sim}1{-}10$~ns, the emission of photons occurs from thermally equilibrated excited
states. This apparent time-scale separation allows us to consider the above
system as a generalized superstatistical system. Therefore, we can find the
limiting probability distribution of the number of ground-mode photons by
directly applying the hierarchical maximum entropy principle to this system.

For simplicity, consider the case of the ground-mode coupling and neglect the
twofold polarization degeneracy by analogy with Ref.~\cite{KlaersEtal2012}. The
whole system is then composed of two subsystems: the subsystem of the dye
solution and the subsystem of the photon BEC. The control parameter
characterizing the interaction of the subsystems is the fluctuating number of
ground-mode photons, $n$. The subsystem of the dye solution in turn consists of
$M$ dye molecules, among which there are $X-n$ excited molecules and $M-X+n$
ground-state molecules. Obviously, $0\leqslant n\leqslant X\leqslant M$. Each
molecule is in contact with a solvent, which plays the role of thermostat. In
this sense, dye molecules resemble cells, but the inverse temperature $\beta$
does not fluctuate. For~$f(\beta)$, this formally corresponds to the conditions
of normalization, a given mean, and zero variance. In what follows, we will not
explicitly indicate the dependence of functions on~$\beta$.

Let $D_0(\varepsilon_0)$ and $D_1(\varepsilon_1)$ be the density
of rovibrational states for the ground, $S_0$, and first excited, $S_1$, singlet
electronic state, respectively. Note that $\varepsilon_i=E-E_i$, where $E_i$ is the
lowest-energy substate of $S_i$, where $i=0,1$. Hence, $D_i(\varepsilon)=0$ for any
$\varepsilon<0$. The partition functions $Z_0$ and $Z_1$ corresponding,
respectively, to the ground-state and excited dye molecules are
\begin{equation}
\label{ZiBeta}
Z_i=e^{-\beta E_i}w_i,
\end{equation}
where
\[
w_i=\int_0^\infty e^{-\beta\varepsilon}D_i(\varepsilon)d\varepsilon,
\qquad
i=0,1.
\]
It follows from Eqs.~\eqref{entropyInEachCell}--\eqref{nu1}
and~\eqref{ZiBeta} that the entropy for a ground-state
molecule, $s_0$, and for an excited molecule, $s_1$, is
\[
s_i=\ln w_i+\beta(u_i-E_i),
\]
where
\begin{equation}
\label{moleculeEnergy}
u_i=E_i-\frac{1}{w_i}\frac{d w_i}{d\beta}
\end{equation}
is the corresponding mean energy.

Now consider the subsystem of all dye molecules. After enumerating them and
denoting a ground-state molecule by~$0$ and an excited molecule by~$1$, we can
write an $M$-digit binary number $\eta=(\eta_1\eta_2\ldots\eta_M)$ with $M-X+n$
zeros and $X-n$ unities such that the state of the $k$th dye molecule is given
by the $k$th digit $\eta_k$. For any given~$\eta$, the entropy of the
corresponding combination of dye molecules is
\[
s_{\eta|n}=(M-X+n)s_0+(X-n)s_1.
\]
The probability that $\eta$ takes on a fixed value is
\[
p_{\eta|n}=%
\begin{pmatrix}
M\\X-n
\end{pmatrix}
^{-1}
=\frac{(X-n)!(M-X+n)!}{M!}.
\]
The entropy $s^\mathrm{d}_n$ of the subsystem of dye molecules is calculated
using the discrete
analogs of Eqs.~\eqref{subsystemEntropy} and~\eqref{betaEntropy}, with
$S[E](\beta|\xi)$ and $f(\beta|\xi)$ replaced by $s_{\eta|n}$ and $p_{\eta|n}$,
respectively:
\[
s^\mathrm{d}_n=s_{\eta|n}+\ln%
\begin{pmatrix}
M\\X-n
\end{pmatrix}
.
\]
The mean energy of the subsystem is
\[
u^\mathrm{d}_n=(M-X+n)u_0+(X-n)u_1,
\]
where $u_0$ and $u_1$ are defined by Eq.~\eqref{moleculeEnergy}.

The entropy of the photon BEC is zero, $s^\mathrm{ph}_n=0$, since
the absence of the polarization degeneracy is
assumed. The total energy of the condensate is
\[
u^\mathrm{ph}_n=n\hbar\omega,
\]
where $\hbar\omega$ is the energy of a ground-mode photon.

Finally, consider the system as a whole. The control parameter $n$ corresponding
to the number of ground-mode photons is characterized by a normalized discrete
probability distribution~$(\pi_0,\ldots,\pi_X)$, where $\pi_n$ is the
probability of $n$ photons. For a fixed $n$, the energy and entropy of the
system are given by $U_n=u^\mathrm{d}_n+u^\mathrm{ph}_n$ and
$S_n=s^\mathrm{d}_n+s^\mathrm{ph}_n$, respectively. Maximizing the entropy
[see~Eqs.~\eqref{totalEntropy} and~\eqref{controlParameterDistributionEntropy}]
\[
S=-\sum_{n=0}^X\pi_n\ln\pi_n+\sum_{n=0}^X\pi_n S_n,
\]
under the normalization condition
\begin{equation}
\label{piNorm}
\sum_{n=0}^X\pi_n=1
\end{equation}
and the mean energy constraint $\sum\pi_n U_n=U$ yields
\begin{eqnarray}
\label{piN}
\pi_n&=&\frac{1}{Z}
\begin{pmatrix}
M\\X-n
\end{pmatrix}
w_0^{M-X+n}w_1^{X-n}\\\nonumber
& &\times\exp\{-\beta[(M-X+n)E_0+(X-n)E_1+n\hbar\omega]\},
\end{eqnarray}
where $Z$ is determined from Eq.~\eqref{piNorm}. Dividing Eq.~\eqref{piN}
by~$\pi_0$ and writing $\hbar\omega_0=E_1-E_0$, we obtain the probability
distribution of the number of ground-mode photons in the form
\begin{equation}
\label{mainBECeq}
\frac{\pi_n}{\pi_0}=\frac{X!(M-X)!}{(X-n)!(M-X+n)!}
\biggl(\frac{w_0}{w_1}\biggr)^n
e^{-\beta n\hbar(\omega-\omega_0)}.
\end{equation}
This equation allows us to find $\pi_0=(\sum\pi_n/\pi_0)^{-1}$ and then
calculate $\pi_n$ for all positive $n\leqslant X$.

Thus, the long-run behavior of the photon BEC, when the probability distribution
$(\pi_0,\ldots,\pi_X)$ becomes stationary, can be investigated using the
hierarchical maximum entropy principle. The link with the result of the master
equation approach can be readily observed via the Kennard-Stepanov law
\cite{Kennard1918,Stepanov1957,Neporent1958,McCumber1964,
SawickiKnox1996,KlaersEtal2012},
\begin{equation}
\label{KennardStepanovLaw}
\frac{B_{10}(\omega)}{B_{01}(\omega)}
=\frac{w_0}{w_1}e^{-\beta\hbar(\omega-\omega_0)},
\end{equation}
which relates the Einstein coefficients for stimulated emission,
$B_{10}(\omega)$, and absorption, $B_{01}(\omega)$.
Equation~\eqref{KennardStepanovLaw} allows us to rewrite Eq.~\eqref{mainBECeq}
as
\[
\frac{\pi_n}{\pi_0}=\frac{X!(M-X)!}{(X-n)!(M-X+n)!}
\biggl[\frac{B_{10}(\omega)}{B_{01}(\omega)}\biggr]^n,
\]
which is identical to Eq.~(10) of Ref.~\cite{KlaersEtal2012}.

It seems interesting to use the described approach for studying the photon BEC
fluctuations in more detail, e.g., for considering a more realistic situation of
the polarization degeneracy and additional fluctuations of~$M$ and~$X$.

\section{Conclusion}

I have formulated the hierarchical maximum entropy principle for generalized
superstatistical systems. Such systems comprise a set of nonequilibrium
superstatistical subsystems, where each subsystem is made up of many cells, and are
characterized by the three-level dynamical hierarchy formed as a result of
the sufficient time-scale separation between different dynamical levels.
By arranging these levels in increasing order of dynamical time scale and
consecutively maximizing the entropy at each level,
I have obtained first the Gibbs canonical
distribution for each cell, second the intensive parameter distribution for
each subsystem, and finally the control parameter distribution for the whole
system. From these distributions, I have also found the superstatistical
distribution for each subsystem and the generalized superstatistical
distribution for the whole system.

I have applied this principle to Bose-Einstein condensation of light in a dye
microcavity. Assuming the ground-mode coupling and neglecting the polarization
degeneracy, I have obtained the long-run probability distribution of the
fluctuating number of ground-mode photons. This distribution is consistent
with the analogous result of the master equation approach.

Note that when the hierarchical maximum entropy principle is applied to a
generalized superstatistical system, certain constraints should be imposed on a
normalized distribution to obtain the canonical distribution at the lower
dynamical level. However, the constraints imposed on the intensive and control
parameter distributions may be quite general. I propose erasing such a
distinction, viz., choosing some general constraints at the lower dynamical
level and additionally considering a vector intensive parameter.
This will result in the generalized superstatistics the local dynamics of
which is described by a more general statistics than the usual Boltzmann-Gibbs
statistics. Grand canonical statistics may be the simplest alternative.

\providecommand{\noopsort}[1]{}\providecommand{\singleletter}[1]{#1}%
\end{document}